\documentclass[a4paper]{jpconf}
\usepackage{graphicx}

\newcommand{\beq}{\begin{equation}}
\newcommand{\eeq}{\end{equation}}
\newcommand{\bqa}{\begin{eqnarray}}
\newcommand{\eqa}{\end{eqnarray}}
\newcommand{\be}{\begin{equation}}
\newcommand{\ee}{\end{equation}}
\newcommand{\bea}{\begin{eqnarray}}
\newcommand{\eea}{\end{eqnarray}}

\begin{document}
\title{Unstable dynamics of
Yang-Mills fields at early times of
heavy ion collisions}

\author{Andreas Ipp}

\address{Institut f\"ur Theoretische Physik, Technische Universit\"at Wien,
Wiedner Hauptstra\ss e 8-10, A-1040 Vienna, Austria}

\ead{ipp@hep.itp.tuwien.ac.at}

\begin{abstract}
The quark gluon plasma as produced in heavy ion collisions is exposed
to early anisotropies in momentum space due to its rapid expansion.
Such anisotropies can lead to non-abelian plasma instabilities,
driven by unstable gluonic modes that can grow exponentially fast.
These plasma instabilities can be simulated using a discretized version
of gauge-covariant Boltzmann-Vlasov and Yang-Mills equations.
In the stationary case, a turbulent cascade forms in the strong-field regime,
which is associated with an approximately linear growth of energy in collective
fields.
Early longitudinal expansion slows down the growth of unstable modes,
and the formation of soft gluonic fields depends crucially
on the initial conditions assumed.
\end{abstract}

\section{Introduction}

Heavy ion colliders like RHIC or LHC create the quark-gluon plasma (QGP)
in an anisotropic state.
Due to its fast expansion, initially along the longitudinal but later also along the transverse directions,
the plasma cools quickly and only exists for a duration of about a few tens of yoctoseconds
(1 ys = $10^{-24}\,\mathrm{s}$).
At early times and close to the center of the QGP,
longitudinal expansion dominates over transverse expansion.
This quickly leads to strong momentum anisotropies along the polar angle
with respect to the collision axis~\cite{Blaizot:2011xf}.
Such early polar momentum space anisotropies
could allow for a violation of the viscosity bound~\cite{Rebhan:2011vd},
or lead to the emission of photon double pulses that are separated merely by yoctoseconds~\cite{Ipp:2009ja}.
Most notably, polar momentum space anisotropies can induce
Chromo-Weibel plasma instabilities,
which are generalizations of Weibel or filamentation instabilities
in ordinary electromagnetic plasmas \cite{Weibel:1959zz}.
It has been suggested early 
that these instabilities may play a fundamental role in the QGP \cite{Mrowczynski:1988dz,Mrowczynski:1993qm,Pokrovsky:1988bm}.
In fact, already an infinitesimal amount of momentum space anisotropy causes the appearance of instabilities in collisionless plasmas \cite{Romatschke:2003ms,Romatschke:2004jh}.
In electromagnetic plasmas, the current filamentation instability 
develops magnetic islands on a
fast electron time scale
induced by the deflection of the original electron streams,
which eventually leads to
an isotropization of the electron distribution \cite{Califano:2001}.
This fast isotropization on a time scale which is faster than ordinary perturbative scattering processes could explain the fast apparent
isotropization and thermalization that is suggested by
hydrodynamic modeling of the early QGP evolution \cite{Arnold:2003rq,Arnold:2004ti}.


\begin{figure}[h]
\begin{minipage}{18.2pc}
\includegraphics[width=18.8pc]{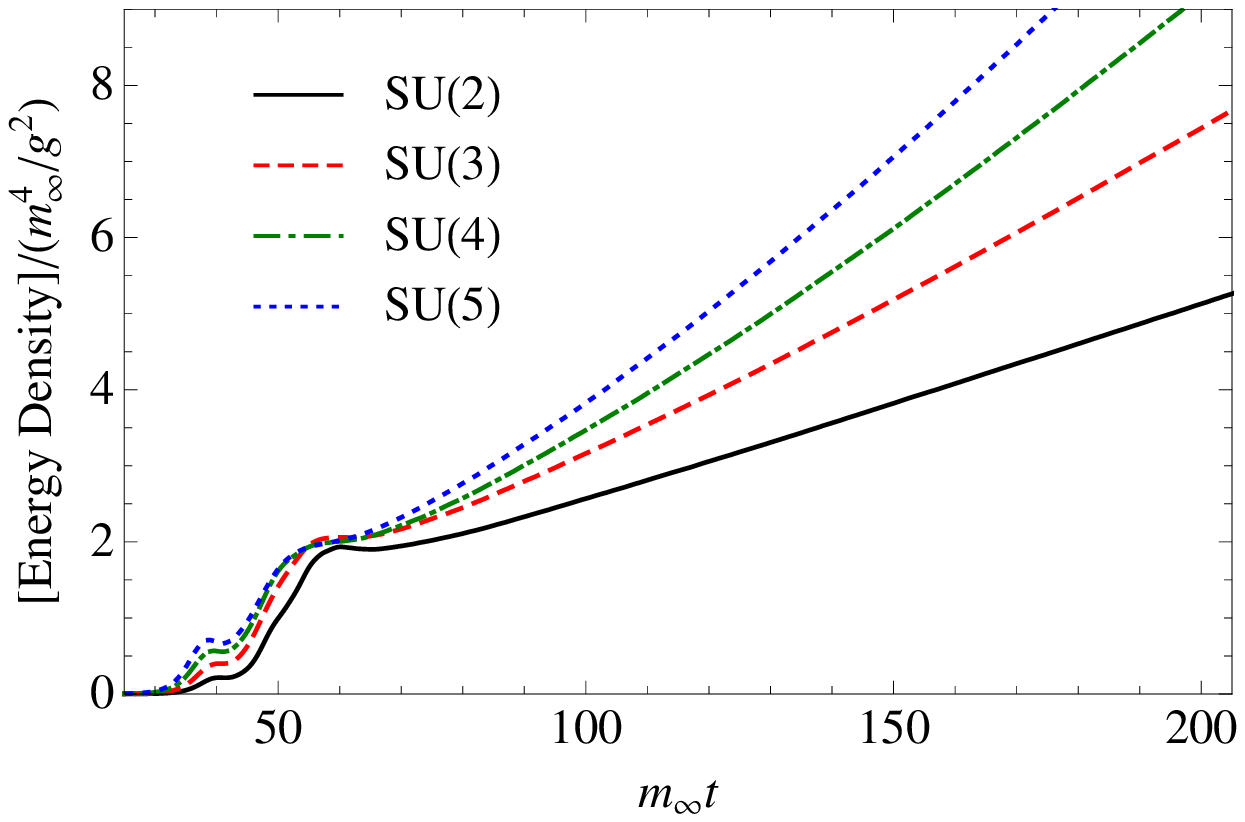}
\caption{Comparison of average 
total field energy densities $\mathcal{E}$ for SU(2) through SU(5) on 
linear scale in
3+1 dimensional simulations for a stationary system with anisotropy parameter $\xi=10$ \cite{Ipp:2010uy}.
\label{fig:energy1}}
\end{minipage}\hspace{1.5pc}%
\begin{minipage}{18.2pc}
\includegraphics[width=19pc]{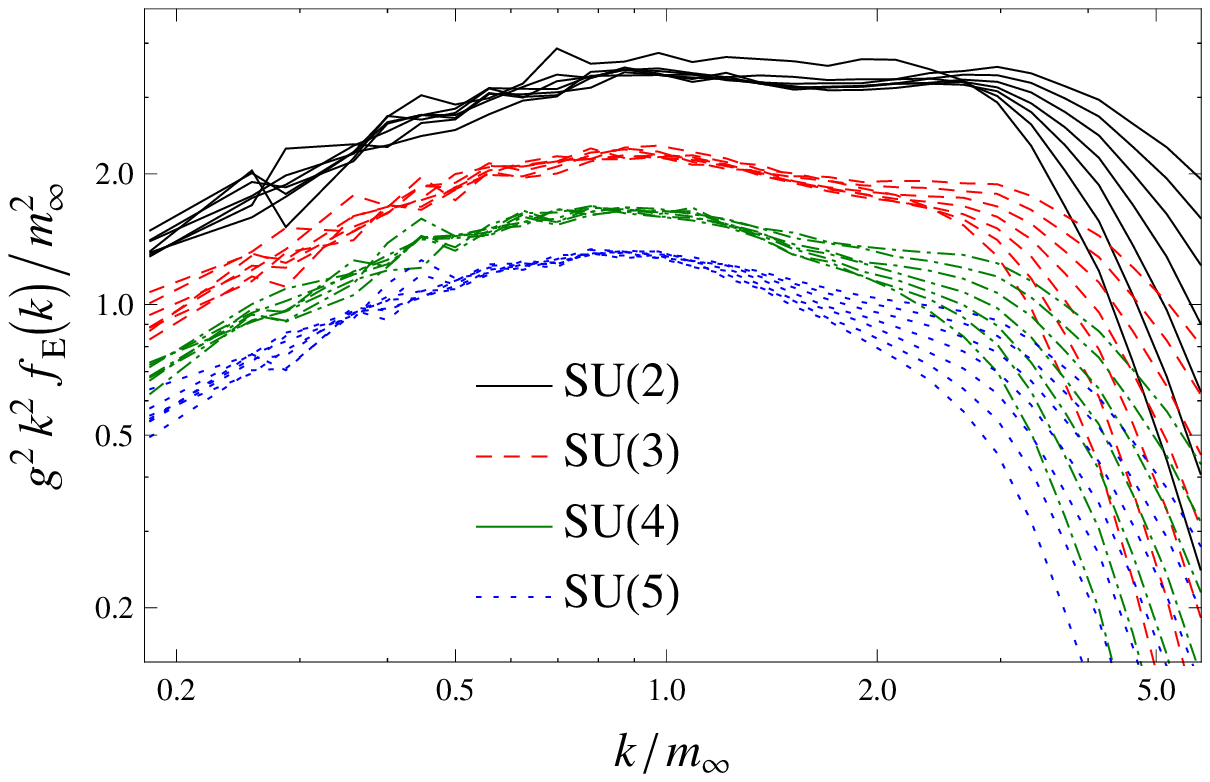}
\caption{The power spectrum for the electric distribution $f_E(k)$
for various gauge groups SU(2) through SU(5)
at late times $80 < m_\infty t < 150$.
The distance between the lines is $m_\infty \Delta t \approx 11$
\cite{Ipp:2010uy}.
\label{fig:spectra2}}
\end{minipage} 
\end{figure}

Non-abelian plasma instabilities can be studied using real-time lattice simulations
\cite{Arnold:2005vb,Rebhan:2005re,
Bodeker:2007fw
}.
The effective field theory for the collective phenomena
at the soft scales is provided by gauge-covariant
collisionless Boltzmann-Vlasov equations 
\cite{Blaizot:2001nr}.
The corresponding effective action is nonlocal and nonlinear
\cite{Pisarski:1997cp,Mrowczynski:2004kv},
but can be made local
using auxiliary fields
in the adjoint
representation $W_\beta(x;\mathbf v)$
which encode the fluctuations of the distribution function of
colored hard particles \cite{Blaizot:1993be}.
They depend on a spatial
unit vector which appears in the velocity $v^\mu=p^\mu/|\mathbf p|$
of a hard particle with momentum $p^\mu$.
The Yang-Mills equations are given by
\begin{equation}
\label{Feq} D_\mu(A) F^{\mu\nu}=j^\nu,
\end{equation}
where the current $j^\nu$ is calculated from the auxiliary fields 
\begin{equation}
\label{Jind0} j^\mu[A] = -g^2 \int {d^3p\over(2\pi)^3} 
{1\over2|\mathbf p|} \,p^\mu\, {\partial f(\mathbf p) \over \partial p^\beta} W^\beta(x;\mathbf v) .
\end{equation}
The non-abelian Boltzmann-Vlasov equation for soft fields reduces to
\begin{equation}
\label{Weq} [v\cdot D(A)]W_\beta(x;\mathbf v) = F_{\beta\gamma}(A) v^\gamma ,
\end{equation}
with $D_\mu=\partial_\mu-ig[A_\mu,\cdot]$. The
scale of the hard particles drops out from these equations.
An anisotropic distribution function $f(\mathbf p)$ is obtained by
deforming an isotropic distribution $f_{\rm iso}$ according to
\be
f(\mathbf p) \propto f_{\rm iso}(\mathbf p^2+\xi p_z^2)
\ee
with anisotropy parameter $-1<\xi<\infty$ \cite{Romatschke:2003ms}.
In order to solve these equations numerically,
the 3-dimensional
configuration space is divided in a cubic lattice,
on which a discretized version of the above
non-abelian gauge-covariant Boltzmann-Vlasov equations
is formulated \cite{Rebhan:2005re}.
The unit sphere of velocities for the auxiliary fields is described 
by a discretized set of unit vectors \cite{Rebhan:2004ur,Rebhan:2005re}
or by an expansion in terms of spherical
harmonics \cite{Arnold:2005vb}.

\section{Numerical results}
In the stationary case, the exponential growth of non-abelian plasma instabilities is limited in 3+1 dimensions
by non-abelian self-interactions.
The exponential growth is limited by gluon self-interactions that
are no longer negligible at a certain magnitude of soft fields.
These self-interactions lead to a turbulence cascade
which form a power-law distribution
$f_k \propto k^{-\nu}$ with a spectral index that turns out to be about $\nu \approx 2$.
While simulations are usually based on the gauge group SU(2) \cite{Arnold:2005ef,Bodeker:2007fw,Arnold:2007cg},
it has been confirmed that the same spectral index holds also in the QCD gauge group SU(3) 
as well as in higher gauge groups \cite{Ipp:2010uy}.
The systematics of the scaling with $N_c$
in the non-abelian regime is shown in Fig.~\ref{fig:energy1}
where the gauge groups SU(2) through SU(5) are compared.
One observes that for
different gauge groups the
energy densities cease to grow at approximately the same value.
In the following linear regime, larger gauge groups grow faster than smaller ones.
Figure \ref{fig:spectra2} shows the late-time behavior of spectra for various
gauge groups.
The spectra are multiplied by $k^2$
so that a scaling with $\nu\approx2$ would correspond to a horizontal line.
The slow growth at large momenta corresponds to the linear growth regime of
Fig.~\ref{fig:energy1}.

\begin{figure}[h]
\begin{minipage}{18.2pc}
\includegraphics[width=17pc]{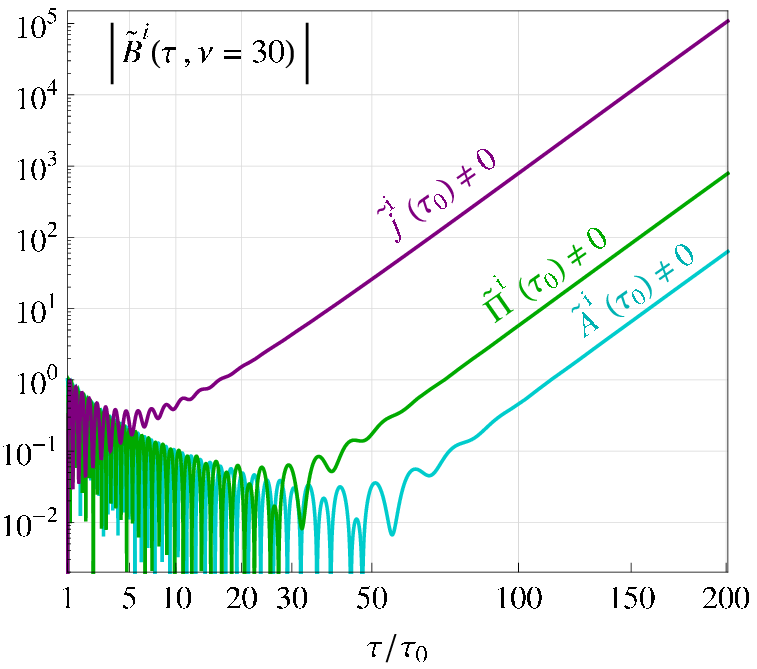}
\caption{Magnetic Weibel instabilities on an expanding background.
The influence of different initial values is studied for modes with wave vector $\nu=30$ in the direction of the anisotropy \cite{Rebhan:2009ku}. 
\label{fig:trans1}}
\end{minipage}\hspace{1.5pc}%
\begin{minipage}{18.2pc}
\includegraphics[width=18.5pc]{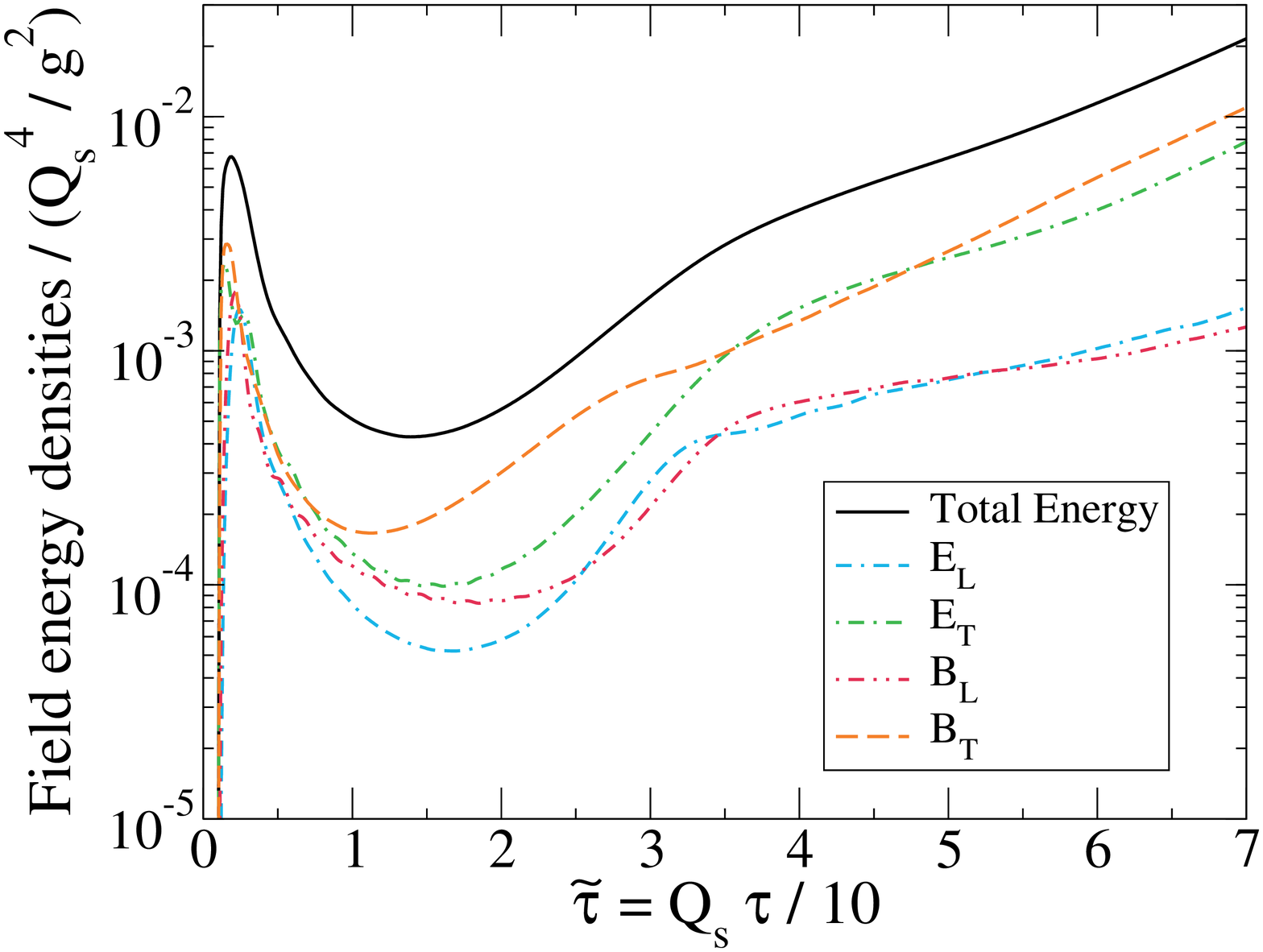}
\caption{Chromoelectric, chromomagnetic, and total energy densities as a function of proper time.
Proper time is normalized such that for $Q_s = 2$ GeV
each unit of $\Delta\tilde\tau$ corresponds to 1 fm/c \cite{Attems:2012js}. 
\label{fig:eDensity1}}
\end{minipage} 
\end{figure}

A longitudinal expansion of the plasma
at early times modifies the exponential growth,
as there are two competing effects:
On the one hand, plasma instabilities drive
the exponential growth, on the other hand
the longitudinal expansion suppresses the growth.
The net effect is a reduction
from a growth exponential in time to exponential in
the square root of proper time.
This has been numerically observed
using the color glass condensate scheme
\cite{Romatschke:2005pm,Romatschke:2006nk}
and the discretized hard loop scheme
\cite{Romatschke:2006wg,Rebhan:2008uj}.
In those simulations,
an uncomfortable delay of the onset of plasma instabilities has been observed \cite{Rebhan:2008uj}.
Collective fields decay and suppress Weibel instabilities at early times of the expanding plasma.
It has been pointed out though that this suppression depends strongly on the initial conditions assumed \cite{Rebhan:2009ku}.
In Fig.~\ref{fig:trans1}, the delay of the growth is displayed for
various possibilities of initial conditions. 
Early work concentrated on a seed electric field with only 
$\tilde{\Pi}^i(\tau_0)\not=0$
\cite{Romatschke:2006wg} 
or a seed magnetic field with only $\tilde{A}^i(\tau_0)\not=0$
\cite{Rebhan:2008uj}.
Seed magnetic fields
instead of seed electric fields increase the delay of plasma instabilities,
as do mixed initial conditions for the fields.
Surprisingly, if one considers initial fluctuations in the
currents with only $\tilde{j}^i(\tau_0)\not=0$,
the delay is very strongly reduced \cite{Rebhan:2009ku}.

This behavior has been confirmed in numerical simulations in 3+1 dimensions. 
Figure \ref{fig:eDensity1} shows various
energy densities as a function of proper time for a longitudinally expanding
system. Initially, the soft fields are depleted by the longitudinal expansion.
Only after some time, the unstable modes overcome the depletion 
and all field components reach approximately the same magnitude.
The growth rate is moderately reduced and transverse chromoelectric and
chromomagnetic fields begin to dominate the
energy density.  
Contrary to the fixed-anisotropy simulations, 
a saturation of the roughly exponential growth is not observed \cite{Attems:2012js}.

Since plasma instabilities may cause the isotropization in plasmas,
the question remains how to measure the isotropization time.
Direct photons would be a good indicator because they leave the QGP likely without further interaction.
Under certain conditions, that is non-central collisions and photon production in a direction close to
forward direction, non-trivial photon pulse shapes may be expected due to intermediate non-isotropic
photon emission \cite{Ipp:2009ja}. In extreme cases, these pulse envelopes may assume the shape
of double pulses at the yoctosecond time scale. 
While it will not be possible to resolve such time structures directly \cite{Ipp:2010vk},
it may be possible to observe the effect of such modifications to the pulse envelope
through Hanbury Brown-Twiss \cite{Ipp:2012zb} correlation measurements.
A photon detector in the required forward direction may be installed during the
ALICE detector upgrade by 2018 when the proposed Forward Calorimeter may be installed.
With a few hundred photon pairs expected per year, such a measurement may be
challenging, but not impossible. Thus, photons could provide valuable information
about the earliest times of the plasma evolution, where gluon dynamics may be subject to plasma instabilities
due to the rapid expansion along the beam axis.

Concluding, one can state that for
heavy-ion collisions at RHIC, 
non-abelian plasma instabilities
probably may not have enough time to develop
as they compete against the fast longitudinal expansion,
but depending on the initial conditions, non-abelian
plasma instabilities may play an important role at LHC energies \cite{Rebhan:2009ku,Attems:2012js}.



\section*{References}

\bibliographystyle{iopart-num}
\bibliography{su3,bibliography}

\end{document}